\documentclass{elsart}%
\usepackage{makeidx}
\usepackage{amsmath}
\usepackage{amsfonts}
\usepackage{amssymb}
\usepackage{graphicx}
\usepackage{lineno}%
\begin{document}
\begin{frontmatter}
\title{Iterative unfolding with the Richardson-Lucy algorithm}
\author{G. Zech\thanksref{mail}}
\thanks[mail]{email: zech@physik.uni-siegen.de}
\address{Universit\"at Siegen, D-57068 Siegen, Germany}
\begin{abstract}
The Richardson-Lucy unfolding approach is reviewed. It is extremely simple 
and excellently performing. It efficiently suppresses artificial high frequency 
contributions and permits to introduce known features of the true distribution. 
An algorithm to optimize the number of iterations has been developed and tested 
with five different types of distributions. The corresponding  unfolding results 
were very satisfactory independent of the number of events, the number of bins 
in the observed and the unfolded distribution, and the experimental resolution.
\end{abstract}
\end{frontmatter}
\linenumbers







\section{Introduction}

In many experiments the measurements are deformed by limited acceptance,
sensitivity or resolution of the detectors. To be able to compare and combine
results from different experiments and to compare the published data to a
theory, the detector effects have to be unfolded. While acceptance losses can
be corrected for, unfolding resolution effects is quite involved. Naive
methods produce oscillations in the unfolded distribution that have to be
suppressed by regularization schemes. 

Various unfolding methods have been proposed in particle physics
\cite{any91,cernworkshop,cowan}. The data are usually treated in form of
histograms. This is also the case in the Richardson-Lucy (R-L) method
\cite{rich72,lucy74} which is especially simple, reliable, independent of the
dimension of the histogram and independent of the underlying metric.

Iterative unfolding with the R-L algorithm has initially been used for picture
restoration. Shepp and Vardi \cite{shepp82,vardi85}, and independently Kondor,
\cite{kondor83} have introduced it into physics. It corresponds to a gradual
unfolding. Starting with a first guess of the smooth true distribution, this
distribution is modified in steps such that the difference between its smeared
version and the observed distribution is reduced. With increasing number of
steps, the iterative procedure develops oscillations. These are avoided by
stopping the iterations as soon as the unfolded distribution, when folded
again, is compatible with the observed data within the uncertainties. We will
discuss the details below. The R-L algorithm originally was derived using
Bayesian arguments \cite{rich72} but it can also be interpreted in a purely
mathematical way \cite{muelthei86,muelthei2005}. It became finally popular in
particle physics after it had been promoted by D'Agostini \cite{dago} with the
label \textquotedblleft Bayesian unfolding\textquotedblright. In Ref.
\cite{lindemann} it was adapted to unbinned unfolding. In Ref. \cite{na38} the
R-L algorithm was applied to a 4-dimensional distribution.

The present situation in particle physics is unsatisfactory for two reasons:
i) There is a lack of comparative systematic studies of the different
unfolding methods and ii) the way to fix the degree of smoothing, the
regularization strength, is usually only vaguely defined.

In the following section we introduce the notation and formulate the
mathematical relations. In Section 3 we discuss regularization and the problem
of assigning errors to the unfolded distribution. In Section 4 the R-L
iterative approach is described. A criterion is developed to fix the number of
iterations that have to be applied and which determine the degree of
regularization. Section 5 contains examples. We conclude with a summary and recommendations.

\section{Definitions and basic relations}

An event sample with variables $\{x_{1},\ldots,x_{n}\}$, the
\emph{input\ sample} is produced according to a statistical distribution
$f(x)$. It is observed in a detector. The \emph{observed sample}
$\{x_{1}^{\prime},\ldots,x_{n^{\prime}}^{\prime}\}$ is distorted due the
finite resolution of the detector and reduced because of acceptance losses. We
distinguish between four different histograms: The \emph{true histogram} with
content $\theta_{j}$, $j=1,\ldots,N$ of bin $j$. $\theta_{j}\propto
\int_{bin\text{ }j}f(x)dx$ corresponds to $f(x)$. The \emph{input histogram}
contains the input sample. The content of its bin $j$ is drawn from a Poisson
distribution with mean value $\theta_{j}$. The \emph{observed histogram}
contains the observed sample with $d_{i}$ events in bin $i$, $i=1,\ldots,M$.
The expected number of events $t_{i}$ in bin $i$ is given by $t_{i}\propto
\int_{bin\text{ }i}f^{\prime}(x^{\prime})dx^{\prime}$ where \ the functions
$f^{\prime}$ and $f$ are related through $f^{\prime}(x^{\prime})=\int
g(x^{\prime},x)f(x)dx$ with the response function $g(x^{\prime},x)$. We choose
$M>N$ \ to constrain the problem. The result of the unfolding procedure is
again a histogram, the \emph{unfolded histogram}, with bin content
$\hat{\theta}_{j}$. We are confronted with a standard inference problem where
the wanted parameters are the bin contents $\theta_{j}$ of the true histogram.
It is to be solved by a least square (LS) or a maximum likelihood (ML) fit. We
discuss only one-dimensional histograms but the corresponding array may
represent a multi-dimensional histogram with arbitrarily numbered cells as well.

The numbers $t_{i}$ and $\theta_{j}$ are related by the linear relation%

\begin{equation}
t_{i}=\sum_{j=1}^{N}A_{ij}\theta_{j} \label{transfer}%
\end{equation}
with the response matrix $A_{ij}$%

\[
A_{ij}=\frac{\int_{bin\text{ }i}f^{\prime}(x^{\prime})dx^{\prime}}%
{\int_{bin\text{ }j}f(x)dx}\;.\;
\]

$A_{ij}$ \ is the probability to observe an event in bin $i$ that belongs to
the true bin $j$. We calculate $A_{ij}$ by a Monte Carlo simulation, but as we
do not know $f(x)$, we have to use a first guess of it. If the size of the 
bins is smaller than the experimental resolution,
the elements of the response matrix show little dependence on the
distribution that is used to generate the events.

We assume that the observed values $d_{i}$ fluctuate according to the Poisson
distribution with the expectation $t_{i}$ and the variance $\delta_{i}%
^{2}=t_{i}$.

The representation of the unfolded distribution by a histogram is a first
smoothing step. We call it \emph{implicit regularization}. With wide enough
bins, strong oscillations in the unfolded histogram are avoided. LS or ML fits
will produce the parameter estimates $\hat{\theta}_{j}$ together with reliable
error estimates. With the prediction $t_{i}$ for $d_{i}$ we can define $\chi^{2}$,

\begin{equation}
\chi^{2}=%
{\displaystyle\sum\limits_{i=1}^{M}}
\frac{\left[  d_{i}-t_{i}\right]  ^{2}}{t_{i}}\;,
\end{equation}
and the log-likelihood $\ln L$ derived from the Poisson distribution,%

\begin{equation}
\ln L=\sum_{i=1}^{M}\left[  d_{i}\ln t_{i}-t_{i}\right]  \;.\label{likstat}%
\end{equation}
Minimizing $\chi^{2}$ or maximizing $\ln L$ determines the estimates of the
parameters $\hat{\theta}_{j}$. The ML fit is applicable also with small event
numbers $d_{i}$ and suppresses negative estimates of the parameter values.
Negative values can occur in rare cases. 

\section{The regularization and the error assignment}

In particle physics the data are often distorted by resolution effects. This means
that without regularization the number of events in neighboring bins of the
unfolded histogram are negatively correlated and as a consequence local
fluctuations are observed. More precisely, the fitted parameters $\hat{\theta}_{j}%
,\hat{\theta}_{j^{\prime}}$ in two true bins $j,j^{\prime}$ are
anti-correlated if their events have sizable probabilities $A_{ij}%
,A_{ij^{\prime}}$ to fall into the same observed bin $i$. These specific
correlations are taken into account in most unfolding methods. An exception is
entropy regularization \cite{nara86,sch94,maga98} which also penalizes
fluctuations between distant bins.

The $\chi^{2}$ surface of the unregularized fit near its minimum $\chi_{0}%
^{2}$ is rather shallow and large correlated parameter changes produce only
small changes $\Delta\chi^{2}$ of $\chi^{2}$ of the fit. The location of the
true parameter point in the parameter space is badly known but the surfaces of
$\chi_{0}^{2}+\Delta\chi^{2}$ for not too small values of $\Delta\chi^{2}$ are
well defined and fix the error intervals which should not be affected by the
regularization. We are allowed to move the point estimate but the error
intervals should not be shifted. The regularization should lead only to a small
increase of $\chi^{2}$. The increase $\Delta\chi^{2}=\chi^{2}-\chi_{0}^{2}$
defines an $N$ dimensional error interval around the fitted point in the
parameter space. It can be converted to a $p$-value%

\begin{equation}
p=\int_{\Delta\chi^{2}}^{\infty}u_{N}(z)dz \label{pvalue}%
\end{equation}
where $u_{N}$ is the $\chi^{2}$ distribution for $N$ degrees of freedom.
Strictly speaking, $p$ is a proper $p$-value only in the limit where the test
quantity $\chi^{2}$ is described by a $\chi^{2}$ distribution. Fixing $p$
fixes the regularization strength. A large value of $p$ corresponds to a weak
regularization and means that the unfolding result is well inside the commonly
used error interval of the likelihood fit. The optimal value of a cut in \ $p$
depends on the unfolding method. Remark that here the value of $\chi_{0}^{2}$
of the fit is irrelevant; what is relevant is its change due to the
regularization. A large value $\chi_{0}^{2}$ could indicate that something is
wrong with the model.

In most applications outside physics, like picture restoration, the
uncertainties of the unfolded distribution are of minor concern. Of interest
are mainly the point estimates which are obtained with a regularization that
the user chooses according to his personal experience. In physics problems,
the error bounds are as important as the point estimates. The manipulations
related to the regularization in most methods constrain the fit and therefore
reduce the errors of the unfolded histogram as provided by the unconstrained
fit \cite{hoecker,truee}. As a consequence, these errors depend on the regularization strength and
do not cover the true distribution with a fixed probability. Distributions
with narrow structures that are compatible with the data may be excluded. An
example for such a behavior is presented in Appendix 1. It is not possible to
associate classical confidence intervals to explicitly regularized solutions.
As stated above, standard error intervals are provided by fits without regularization.

In the iterative method the errors could in principle be calculated by error
propagation but these errors would not be constrained and therefore usually be
large and strongly correlated. Furthermore their interpretation would be
difficult. Therefore it does not make sense to include them in the graphical
representation. A very qualitative way to indicate the errors is presented in
Appendix 2.

To document quantitatively the precision of the data, a fit with a small
number of bins and without explicit regularization of the unfolded histogram
should be done, such that by a wide enough binning artificial oscillations are
sufficiently suppressed. The result together with the corresponding error
matrix\footnote{Instead of the error matrix its inverse could be published. 
The inverse is needed if data are combined or if parameters are estimated.}
estimate contain the information that is necessary
for a comparison with theoretical predictions or other experiments. An example
is given in Appendix 2. Alternatively, the data vector and the response matrix
could be kept. These items, however, require some explanation to non-experts.

In case we have a
theoretical prediction in analytic form, depending on unknown parameters, we
should avoid unfolding and the regularization problem and estimate the
parameters directly \cite{bohm}. A direct fit does not require the 
construction of a response matrix and is independent of assumptions about the shape of the 
distribution used to simulate the experiment, parameter inference is possible 
even with very low event numbers where unfolding is problematic, the 
results are unbiased and the full information contained in the experimental data
can be explored.

\section{The Richardson-Lucy iteration}

\subsection{The method}

Replacing the expected number $t_{i}$ in relation (1) by the observed number
$d_{i}$, the corresponding matrix relation $d=A\hat{\theta}$ can be solved
iteratively for the estimate $\hat{\theta}$. The idea behind the iteration
algorithm is the following: Starting with a preliminary guess $\hat{\theta
}^{(0)}$of $\theta$, the corresponding prediction for the observed
distribution $d^{(0)}$ is computed. It is compared to $d$ and for a bin $i$
the ratio $d_{i}/d_{i}^{(0)}$ is formed which ideally should be equal to one.
To improve the agreement, all true components are scaled proportional to their
contribution $A_{ij}\hat{\theta}_{j}^{(0)}$ to $d_{i}^{(0)}$. This procedure
when iterated corresponds to the following steps:

The prediction $d^{(k)}$ of the iteration $k$ is obtained in a \emph{folding
step} from the true vector $\hat{\theta}^{(k)}$:%

\begin{equation}
d_{i}^{(k)}=\sum_{j=1}^{N}A_{ij}\hat{\theta}_{j}^{(k)}\;. \label{folding}%
\end{equation}

In an \emph{unfolding step}, the components $A_{ij}\hat{\theta}_{j}^{(k)}$ are
scaled with $d_{i}/d_{i}^{(k)}$ and added up into the bin $j$ of the true
distribution from which they originated:%

\begin{equation}
\hat{\theta}_{j}^{(k+1)}=\sum_{i=1}^{M}A_{ij}\hat{\theta}_{j}^{(k)}\frac
{d_{i}}{d_{i}^{(k)}}\left/  \alpha_{j}\right.  \;. \label{unfolding}%
\end{equation}

Dividing by the acceptance $\alpha_{j}=\sum_{i}A_{ij}$ corrects for acceptance losses.

The result of the iteration converges to the maximum likelihood solution as
was proven by Vardi et al. \cite{vardi85} and M\"{u}lthei and Schorr
\cite{muelthei86} for Poisson distributed bin entries. Since we start with a
smooth initial distribution, the artifacts of the unregularized ML estimate
(MLE) occur only after a certain number of iterations.

The regularization is performed simply by interrupting the iteration sequence.
As explained above, the number of applied iterations should be based on a
$p$-value criterion which measures the compatibility of the regularized
unfolding solution with the MLE.

To this end, first the number of iterations is chosen large enough to approach
the asymptotic limit with the ML solution which provides the best estimate of 
the true histogram if we put aside our prejudices about smoothness.
Folding the result and comparing it to the observed histogram, we obtain 
$\chi_{0}^{2}$ of the fit.
\begin{figure}
[ptb]
\begin{center}
\includegraphics[
height=3.4982in,
width=4.4815in
]%
{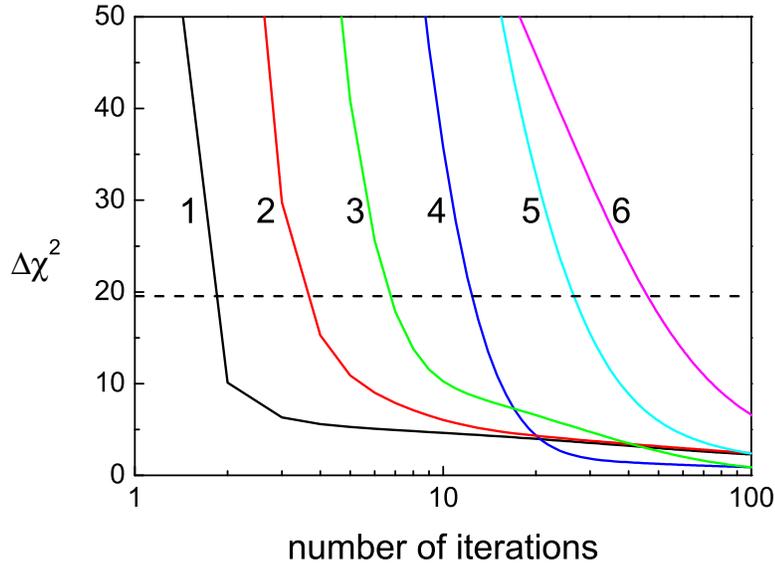}%
\caption{{\small Convergence of the iterative unfolding for distributions with
1 to 6 oscillations.}}%
\label{oscil}%
\end{center}
\end{figure}

Of course, the MLE does not depend on the starting distribution but the
regularized solution obtained by stopping the iteration does. We may choose it
according to our expectation. In most cases the detailed shape of it does not
matter, and a uniform starting distribution will provide reasonable results.

As may be expected, the speed of convergence decreases with the spatial
frequency of the true distribution if we consider a Gaussian type of smearing
described by a point spread function. This is shown in Fig. \ref{oscil}. Here
the true distributions consisting of a superposition of a uniform distribution
of $1000$ events and a squared sine/cosine distributions of $9000$ events with
$1$ to $6$ oscillations is smeared and distributed into $40$ bins. The
corresponding histogram is unfolded to a $20$ bin histogram starting with a
uniform histogram. The statistic $\Delta\chi^{2}$ for $20$ degrees of freedom
is plotted as a function of the number of iterations. The discrete points are
connected by a line. The horizontal line corresponds to a $p$-value of $0.5$. As
expected, the number of required iteration steps that are needed to reach the
$p=0.5$ value increases with the frequency of the distribution. This means
that high frequency contributions and artificial fluctuations of correlated
bins are strongly suppressed in the R-L approach. The reason can be inferred
from Relation (\ref{unfolding}): The parameters $\theta_{j},\theta_{j^{\prime
}}$ of bins $j,j^{\prime}$ that are correlated in that they have
similar\ values $A_{ij}$, $A_{ij^{\prime}}$ are scaled in a similar way and
relative fluctuations develop only slowly with increasing number of iterations.

\emph{Remark}: By construction, the R-L method is invariant against an
arbitrary re-ordering of the bins. A multidimensional histogram can be
transformed to a one-dimensional histogram. A rather general class of
distortions can be treated. This is also true for entropy regularization and
methods based on truncation of the eigenvalue sequence in singular value
decomposition (SVD) \cite{hoecker} but not for local regularization schemes
like curvature suppression \cite{tikhonov} which is difficult to apply in
higher dimensions.%

\begin{figure}
[ptb]
\begin{center}
\includegraphics[
height=4.8447in,
width=6.5406in
]%
{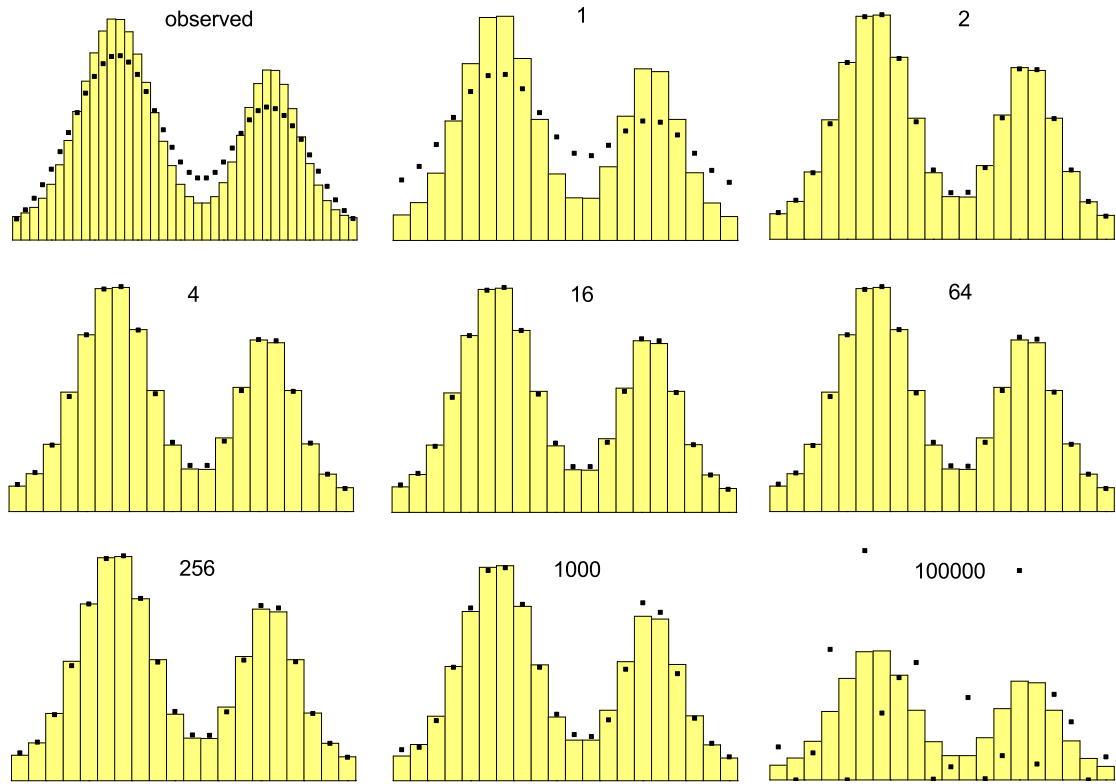}%
\caption{{\small Observed histogram and unfolded histograms (squares) for
different number of iterations compared to the input histogram (shaded).}}%
\label{iter2b2040}%
\end{center}
\end{figure}

\subsection{The regularization strength}

Without recipes how to fix the regularization strength, unfolding methods are
incomplete and the results are to a certain extent arbitrary. In most of the
proposed methods a recommendation is missing or rather vague. In the iterative
method, we have to find a criterion, based on a $p$-value, when to stop the
iteration process. The optimum way may depend on several parameters: the
number of events, the number of bins, the resolution and the shape of the true
distribution. Not all combinations of these parameters can be investigated in
detail. We will study some specific Monte Carlo examples to derive a stopping
criterion and then test it with further distributions. It will be shown that a
general prescription works reasonably well for all studied examples.

The unfolded histogram is compared to the input histogram. In all examples we
take care that the estimates of the elements of the response matrix have
negligible statistical uncertainties. If not stated differently, the iteration
starts with a uniform distribution as a first guess for the true distribution.
The observed histogram has, with two exceptions, $40$ bins and the unfolded histogram usually
comprises $20$ bins. With the standard settings the value of $\chi_{0}^{2}$ should be compatible with the
$\chi^{2}$ distribution with $20$ degrees of freedom because we have $40$
measurements and $20$ estimated parameters.

\subsubsection*{Example 1: Two peaks}

We start with a two-peak distribution, a superposition of two normal
distributions $N(x|0.3;0.10)$, $N(x|0.75;0.08)$ and a uniform distribution
$U(x)$ in the interval $0<x<1$. Here $N(x|\mu;\sigma$) is the normal
distribution of $x$ with the mean value $\mu$ and the standard deviation
$\sigma$. The number of events attributed to the three distributions is
$25,000$, $15,000$ and $10,000$, respectively. The experimental distribution is
observed with a Gaussian resolution $\sigma=0.07.$ It is of the same order as
the width of the peaks. Events are accepted in the interval $0<x,x^{\prime}<1$.

In Fig. \ref{iter2b2040} unfolding results for different values of the number
of iterations are shown. The shaded histograms (input histograms) correspond
to the observation with an ideal detector and are close to the true histogram.
The left top plot displays the observed histogram as squares. With increasing
number of iterations the unfolded histogram (squares) quickly approaches the
true histogram. The agreement is quite good in a wide range of the number of
iterations. It deteriorates slowly when increasing the number of iterations
beyond $32$. At $1000$ iterations oscillations are visible and after $100,000$
iterations the sequence has approached the maximum likelihood solution with
strong fluctuations and no explicit regularization. We find $\chi_{0}%
^{2}=23.4$ for $20$ degrees of freedom.%

\begin{figure}
[ptb]
\begin{center}
\includegraphics[
height=4.6417in,
width=6.237in
]%
{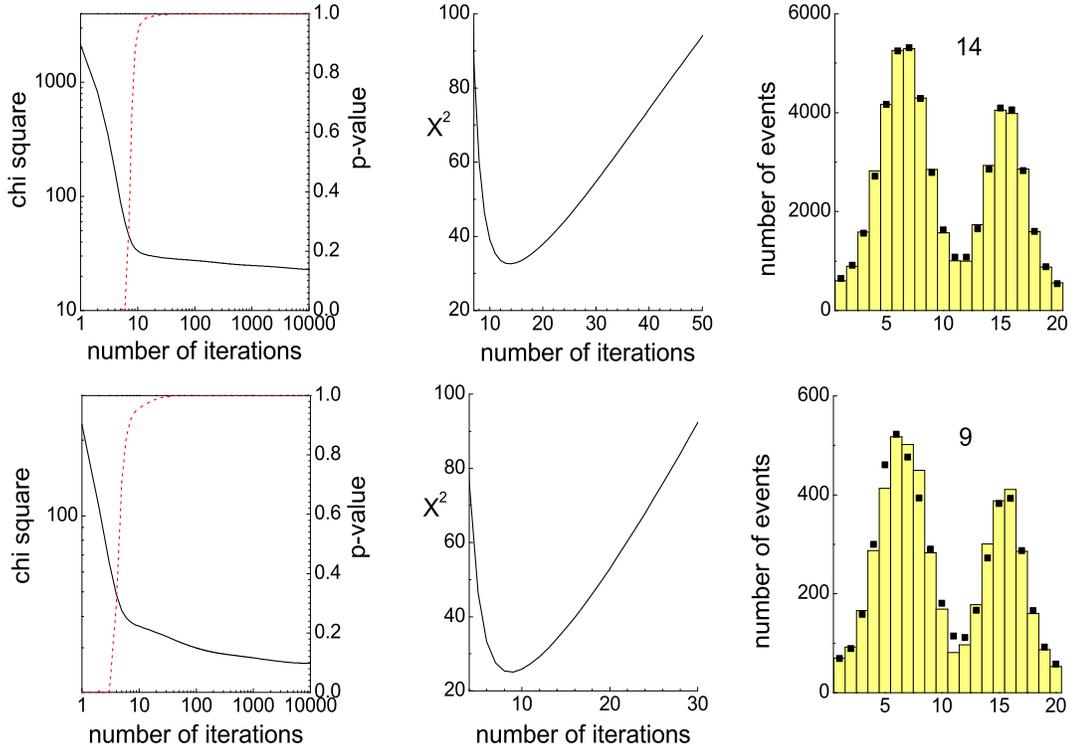}%
\caption{$\chi^{2}${\small \ and $p$-value as a function of the number of
iterations (left), }$X^{2}${\small \ as a function of the number of iterations
(center) and unfolding result after the optimal number of iterations (right).
The upper plots correspond to }$50,000${\small \ the lower ones to }%
$5,000${\small \ events.}}%
\label{bumbprob}%
\end{center}
\end{figure}

The variation of $\chi^{2}$ as a function of the number of iterations is shown
in Fig. \ref{bumbprob} top, left hand scale. The corresponding $p$-value (right
hand scale) jumps within a few iterations from a negligible value to a value
close to one. To judge the quality of the unfolding, we compute the quantity
$X^{2}=\Sigma_{i}(\hat{\theta}_{i}-\theta_{i})^{2}/\theta_{i}$ which is
available in toy experiments. It is difficult to estimate the range of values
of $X^{2}$ that correspond to acceptable solutions, but qualitatively the
agreement of the unfolded histogram with the true histogram improves with
decreasing $X^{2}$. The dependence of $X^{2}$ from the iteration number is
displayed at the top center of the same figure. The minimum is reached at
$14$ iterations with a $p$-value of $0.98$ but there is little change
between $8$ and $16$ iterations. The corresponding unfolding result is shown
on the right hand side. Repeating the same experiment with ten times less
events, i.e. $5,000$, we obtain the results displayed at the bottom of Fig.
\ref{bumbprob}. Here the best agreement is reached after $9$ iterations.%

\begin{figure}
[ptb]
\begin{center}
\includegraphics[
height=3.2471in,
width=5.7103in
]%
{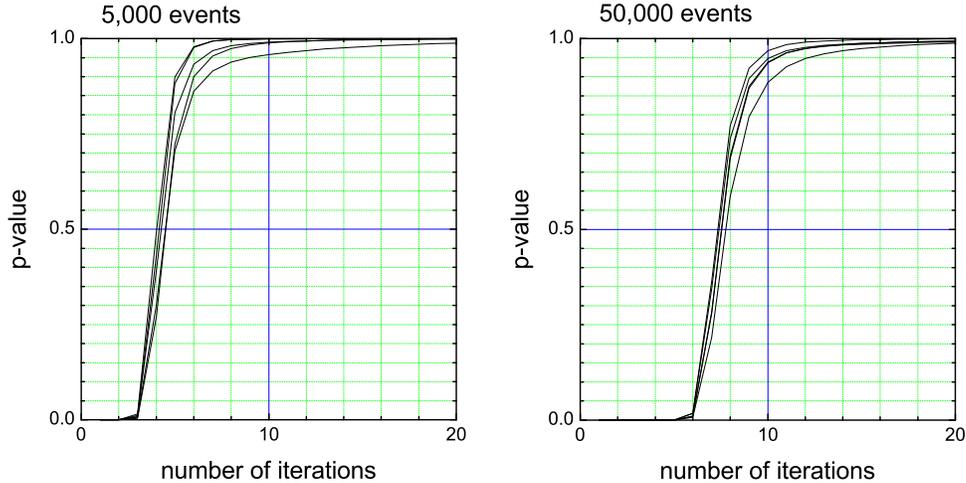}%
\caption{{\small $P$-value as a function of the number of iterations.}}%
\label{probcut}%
\end{center}
\end{figure}

The study is repeated for $5$ different samples. The $p$-values are shown as a
function of the number of iterations in Fig. \ref{probcut}. All curves start
rising nearly at the same iteration, remain close to each other at the
beginning but separate at large $p$-values. With $5,000$ events the lowest value
of the test quantity $X^{2}$ is always obtained for $8$ or $9$ iterations,
while the corresponding $p$-values vary because of the small slopes near
$p$-values of one. Therefore, we should base the cut of the chosen number of
iterations on a lower $p$-value. The following choice has proven to be quite
stable and efficient: We stop the iteration at twice the value at which the
$p$-value crosses the $0.5$ line. For the left hand plot with $5,000$ events the
crossing is close $4.5$ and thus $9$ iterations should be performed. With
$50,000$ events this criterion leads to a choice of $15$ iterations. Actually,
from the $X^{2}$ variation, acceptable values are located between $11$ and
$16$ iterations. In Table 1 the results for the same distribution but
different number of bins of the observed and the unfolded histogram and for different
resolutions $\sigma$ are summarized. From left to right the columns contain the
number of generated events, the number of bins in the observed and the true
histograms, the standard deviation of the Gaussian response function, the
number of applied iterations as based on the stopping criterion, $X^{2}$, the
corresponding $p$-value, the number of iterations that minimizes $X^{2}$ and the
minimal value of $X^{2}$. In each case two independent toy experiments have
been performed. The results from the second one are given in parentheses. They
are close to those of the first one. In all cases the recipe for the choice of
the number of iterations leads in most cases to very sensible results. The
$p$-values are close to $1$ in most cases and always above $0.95$.

For the resolution $0.1$ the optimal number of iterations and also the $X^{2}$
values differ considerably from the those found by the stopping criterion. The
visual inspection shows however that the unfolded distributions that
correspond to the stopping prescription agree qualitatively well with the true
distributions. For comparison, the example with $50,000$ events and resolution
$0.1$ has also been repeated with a likelihood fit and entropy penalty regularization. The
regularization constant was varied until the minimum
of $X^{2}$ was obtained. The results was $X^{2}=159$ significantly larger
than the value $91$ obtained with iterative unfolding. 
With the prescription $\Delta\chi^2=1$ \cite{sch94}, $X^{2}=873$ was obtained.
Regularization with a curvature penalty is not suited for this example. Here the best value of $X^2$ is $700$.

\begin{table}[ptb]
\caption{Test of the $p$-value cut}%
\label{testpv}%
\centering
\par%
\begin{tabular}
[c]{|l|l|l|l|l|l|l|l|}\hline
events & bins & $\sigma$ & $\#$ & $X^{2}$ & $p$-value & $\#_{best}$ &
$X_{best}^{2}$\\\hline
50000 & 40/20 & 0.07 & 15 (15) & 33 (40) & 0.989 (0.986) & 15 (14) & 33 (40)\\
5000 & 40/20 & 0.07 & 9 (8) & 25 (39) & 0.958 (0.980) & 9 (9) & 25 (39)\\
50000 & 40/14 & 0.07 & 18 (16) & 25(32) & 0.978 (0.989) & 16 (17) & 25 (32)\\
5000 & 40/14 & 0.07 & 9 (10) & 27 (40) & 0.997 (0.971) & 10 (8) & 26 (38)\\
50000 & 40/30 & 0.07 & 13 (13) & 44 (45) & 1.000 (1.000) & 14 (15) & 44 (44)\\
5000 & 40/30 & 0.07 & 7 (7) & 28 (39) & 0.997 (1.000) & 8 (8) & 27 (39)\\
50000 & 40/20 & 0.05 & 8 (8) & 31 (21) & 1.000 (1.000) & 7 (11) & 31 (21)\\
5000 & 40/20 & 0.05 & 5 (6) & 9 (22) & 0.997 (0.971) & 6 (5) & 9 (20)\\
50000 & 40/20 & 0.10 & 33 (33) & 143 (148) & 1.000 (1.000) & 205 (176) & 91
(108)\\
5000 & 40/20 & 0.10 & 15 (18) & 100 (57) & 1.000 (0.985) & 23 (23) & 77 (52)\\
50000 & 80/20 & 0.7 & 15 (15) & 32 (37) & 0.991 (0.985) & 14 (15) & 32 (37)\\
5000 & 80/20 & 0.7 & 8 (8) & 26 (36) & 0.970 (0.999) & 7 (8) &26
(36)\\\hline
\end{tabular}
\end{table}

\subsubsection{Interpolation for fast converging iterations}

In situations where the response function is narrow, usually the iteration
sequence converges quickly to a reasonable unfolded histogram, sometimes after
a single iteration. Then one might want to stop the sequence somewhere between
two iterations. This is possible with a modified unfolding function. We just
have to introduce a parameter $\beta>0$ into (\ref{unfolding}):%

\begin{equation}
\hat{\theta}_{j}^{(k+1)}=\left[  \sum_{i=1}^{M}A_{ij}\hat{\theta}_{j}%
^{(k)}\frac{\hat{d}_{i}}{d_{i}^{(k)}}\left/  \alpha_{j}\right.  +\beta
\hat{\theta}_{j}^{(k)}\right]  \;\left/  (1+\beta)\right.  .
\label{unfoldingsmooth}%
\end{equation}

The value $\beta=0$ produces the original sequence (\ref{unfolding}), with
$\beta=1$ the convergence is slowed down by about a factor of two and in the
limit where $\beta$ approaches infinity, there is no change. It is proposed to
choose $\beta$ such that at least $5$ iteration steps are performed.%

\begin{figure}
[ptb]
\begin{center}
\includegraphics[
trim=0.124146in 0.103943in 0.049890in 0.000000in,
height=8.2399in,
width=6.5743in
]%
{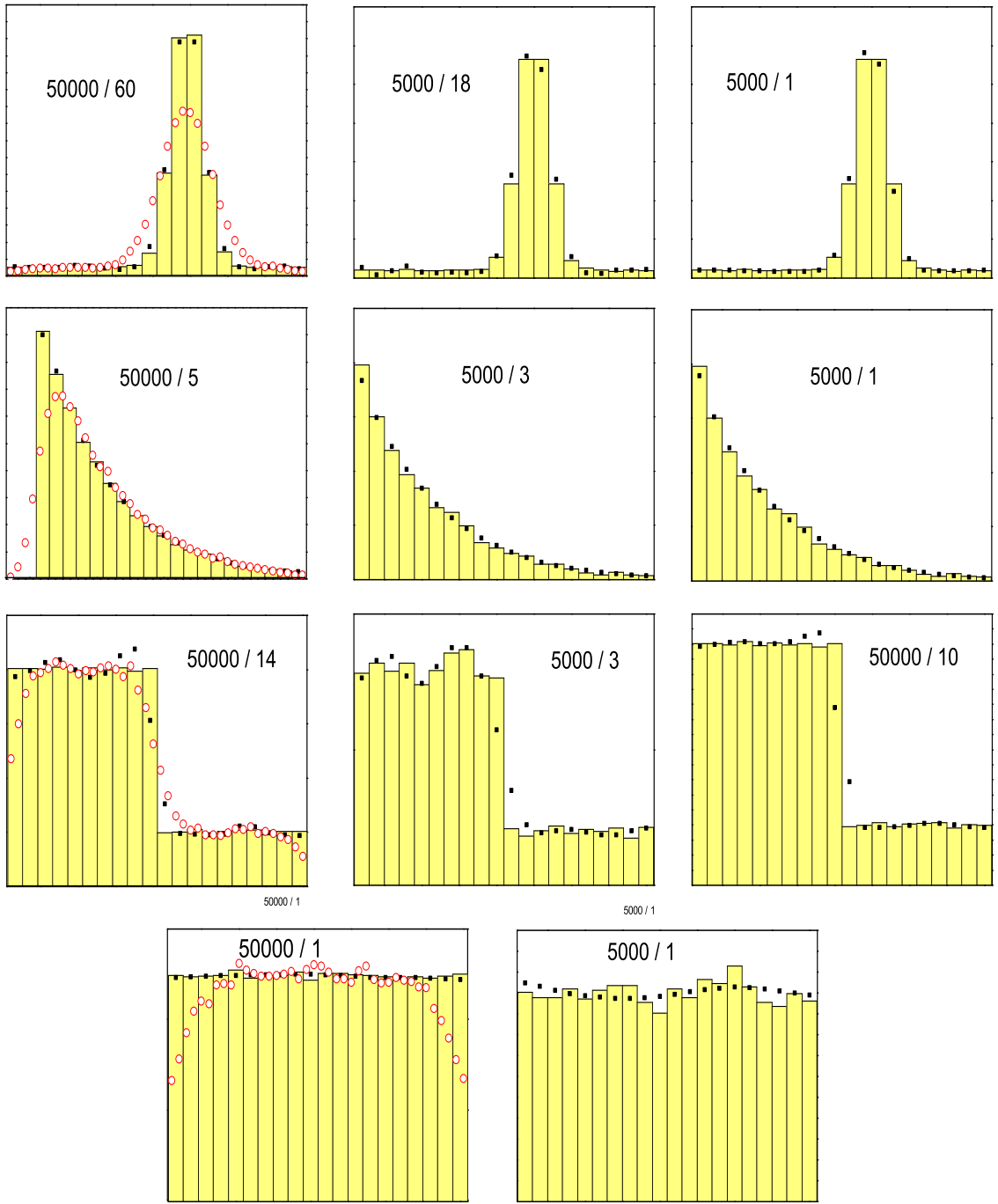}%
\caption{{\small Unfolding results (squares) for different distributions. The
shaded histogram corresponds to the input to the detector, the hollow circles
to the observed histogram. The number of generated events and the number of
applied iterations are indicated.}}%
\label{iterpub}%
\end{center}
\end{figure}

\subsection{Subjective elements}

Unfolding is not an entirely objective procedure. The choice of the method and
the kind of regularization depend at least partially on personal taste. For a
given value of $\chi^{2}$ there exist an infinite number of unfolded
histograms. There is no objective criterion which would allow us to choose the
best solution. Given the R-L iterative unfolding, with the stopping criterion
as defined above and a uniform starting distribution all parameters are fixed,
but in some rare situations it may make sense to modify the standard method.

\subsubsection{Choice of the starting distribution}

Instead of a uniform histogram we may choose a different starting histogram.
As long as the corresponding distribution shows little structure, the
unfolding result will not be affected very much. If we start in our Example 1
($50,000$ events) with an exponential distribution $f(x)=e^{-x}$ the unfolded
histogram is hardly distinguishable from that with a uniform starting
distribution. The difference is less than \ $1\%$ in all bins except for the
two border bins with only about $60$ entries where it amounts to $2\%$. In
both cases $15$ iterations are required.

For an input distribution that is close to the true distribution, the results
are in most cases again very similar to those of the uniform input
distribution, but of course the number of required iterations is reduced to
one ore two. The situation is different for distribution with sharp
structures, for instance, if there is a narrow peak with a small smooth
background. Starting with a uniform distribution a large number of iterations
is required which may lead to oscillations in the background region. This
unpleasant effect is avoided if we start with a distribution that includes a
peak structure and where only few iterations are necessary.

We have to be careful when choosing a starting distribution different from a
monotone function. Only statistically well established structures should be
modeled in the starting distribution.

The starting distribution can be obtained by fitting a polynomial, spline
functions or another sensible parametrization to the data with the method
described in Ref. \cite{bohm}.

\subsubsection{Manual smoothing}

In the specific example with a narrow peak which we discuss below, starting with a uniform
distribution we can also avoid the oscillations if we replace the oscillating
part in the true input histogram by a smooth distribution before the last
iteration step\footnote{A similar but more drastic proposal has been made in
Ref. \cite{dago1}.}.

\section{Examples with various distributions}

We test the R-L unfolding and the stopping criterion with four different
distributions, a single peak distribution, an exponential distribution, a step
distribution and a uniform distribution. The results are displayed in Fig.
\ref{iterpub}. The number of events and the number of iterations are indicated
in the plots. The starting true function is uniform, except for the last
column where a rough guess of the true distribution is used. The input
histogram is shaded, the unfolded histogram is indicated by squares and the
observed histogram is plotted as circles in the left hand graphs.

\subsubsection*{Example 2: Single narrow peak}

We turn now to a more difficult problem and consider a distribution of
$40,000$ events distributed according to $N(x|0.6;0.05)$ and $10,000$ events
distributed uniformly. The Gaussian response function with $\sigma=0.07$ is
wider than the peak. There is a problem because for the flat region we would
be satisfied with few iterations while the peak region requires many
iterations. Here $\operatorname{about}$ $60$ iterations are needed because
relatively high frequencies are required to model the narrow peak. We get
$\chi^{2}=27$ while the value of $\chi_{0}^{2}$ after $100,000$ iterations is
$20.6$. The unfolded histogram is shown in Fig. \ref{iterpub} top left
together with the smeared histogram and the true histogram. The peak is well
reproduced. The corresponding results for $5,000$ events is shown at the
center of the first row. The right hand plot is obtained with a modified input
distribution for the last iteration. The unfolding result after $18$
iterations is used as input, but the flat region is replaced by a uniform
distribution and one additional iteration is applied. In this way the
artificial oscillations in the background region are reduced.

To test the effect of an improved starting distribution, a superposition of a
quadratic basic spline function (b-spline) and a uniform distribution was
fitted to the data. Four parameters were adjusted, two
normalization\ parameters, the location and the width of the b-spline bump.
With this starting distribution, after a single iteration the input
distribution is almost perfectly reproduced. The test quantity $X^{2}$ is $47$
compared to $216$ with a uniform starting distribution.

\subsubsection*{Example 3: Exponential distribution}

$50,000$ events are generated in the interval $1<x<5$ according to an
exponential distribution $f(x)=e^{-x}$ and $\sqrt{x}$ is smeared with a
Gaussian resolution of $\sigma=0.1$ which means that the smearing of $x$
increases proportional to $\sqrt{x}$. The events are observed in the interval
$0.5<x^{\prime}<5$ and distributed into $40$ bins. The convergence is rather
fast because the distribution is smooth even though we start with a uniform
true distribution. We stop after $5$ iterations and get $\chi^{2}=31.5$ which
corresponds to a $p$-value of $0.996$. The results are shown in the second row
of Fig. \ref{iterpub}. In fact the agreement of the unfolded distribution
improves slightly with additional iterations and is optimum after $7$
iterations. With $5,000$ events the convergence is faster and a reasonable
agreement is obtained after $3$ iterations. Starting with a first guess of an
exponential distribution the result slightly improves (right hand plot).

\subsubsection*{Example 4: Step function}

A step function is rather exotic. The sharp edge is not easy to reconstruct.
We locate the edge at the center of the interval and superpose two uniform
distributions containing $40,000$ events in the interval $0<x<0.5$ and
$10,000$ events in the interval $0.5<x<1$ with the resolution $\sigma=0.05$.
The unfolding results shown in the third row of Fig. \ref{iterpub} are
disappointing. The $p$-value of $\ 0.99$ is reached after $25$ iterations with
$\chi^{2}=20.63$ ($\chi_{0}^{2}=12.42$). A problem is that to model the sharp
edge, many iterations are required while for the flat regions oscillations
start after a few iterations. However if we replace the uniform starting
distribution by the result displayed in the left hand plot replacing the $16$
bins of the flat region by uniform distributions the result (right hand plot)
near the edge is not improved

\subsubsection*{Example 5: Uniform distribution}

A uniform distribution is easy to unfold. $50,000$ events generated in the
interval $0<x<1$ with a Gaussian resolution of $\sigma=0.1$ and observed in
the same interval are unfolded. As the iteration starts with a uniform
distribution, no iteration is necessary and the result is optimal with a
$p$-value close to one. The initial value of $\chi^{2}$ is $26.4$ and the
minimum value is $19.3$ corresponding to the strongly oscillating ML solution.
In the case of $5,000$ events $1$ iteration is applied.

\begin{table}[ptb]
\caption{Test of the stopping criterion}%
\label{teststop}
\centering
\begin{tabular}
[c]{|l|rrr|rrr|rrr|rrr|rrr|}\hline
case & \multicolumn{3}{|c|}{1 peak} & \multicolumn{3}{|c|}{2 peak} &
\multicolumn{3}{|c|}{exponential} & \multicolumn{3}{|c|}{step} &
\multicolumn{3}{|c|}{uniform}\\
& $\chi^{2}$ & $X^{2}$ & \# & $\chi^{2}$ & $X^{2}$ & \# & $\chi^{2}$ & $X^{2}$
& \# & $\chi^{2}$ & $X^{2}$ & \# & $\chi^{2}$ & $X^{2}$ & \#\\\hline
50,000 & 27 & 216 & 60 & 31 & 33 & 15 & 29 & 20 & 10 & 24 & 600 & 14 & 26 &
3 & 0\\
50,000 best & 29 & 209 & 51 & 31 & 33 & 15 & 30 & 19 & 7 & 18 & 488 & 48 &
26 & 3 & 0\\
5,000 & 32 & 167 & 18 & 37 & 25 & 9 & 43 & 7 & 2 & 37 & 104 & 3 & 45 & 6 & 1\\
5,000 best & 29 & 71 & 70 & 37 & 25 & 9 & 43 & 7 & 2 & 33 & 96 & 6 & 45 & 6 &
1\\\hline
\end{tabular}
\end{table}

\subsubsection{Test of the stopping criterion}

In Table 2 we compare the result obtained with the stopping criterion to the
result obtained with the optimal number of iterations (denoted by \emph{best}
in the table). In all cases the iteration starts with a uniform distribution.
The agreement with the observed distribution, indicated by $\chi^{2}$, the
compatibility of the unfolded distribution with the input distribution,
measured with $X^{2}$ and the number of applied iterations are given. The
stopping criterion produces very satisfactory results in all cases. With the
exception of the single peak distribution with $5,000$ events, it is close to
the optimum. Here the observed discrepancy between the number of iterations
from the stopping criterion and the number derived from the minimum of $X^{2}$
is due to the fact that the distribution consists of a flat region where few
iteration are needed and the peak region which requires many iteration to
converge to an optimal result. Nevertheless also the solution with $18$
iteration is satisfactory.

\section{Summary, conclusions and recommendations}

Iterative unfolding with the R-L approach is extremely simple, independent of
the number of dimensions, efficiently damps oscillations of correlated
histogram bins and needs little computing time. A general stopping criterion
has been introduced that fixes the number of iterations, e.g. the
regularization strength, that should be applied. It has a simple statistical
interpretation. Its stability has been demonstrated for five different
distributions, two different event numbers, two different experimental
resolutions and three binnings. The results are very satisfactory. The present
study should be extended to more distributions with varying statistics and
binning and also be applied to higher dimensions.

In most problems a uniform distribution should be used as starting
distribution, but the dependence on its shape is negligible as long as this
distribution does not contain pronounced structures. In cases where the
observed distribution indicates that there are sharp structures in the true
distribution, the iterative method permits to implement these in the input
distribution. In this way the number of iterations is reduced and oscillations
are avoided.

Standard errors, as we associate them commonly in particle physics to
measurements, cannot be attributed to explicitly regularized unfolded
histograms. We propose to indicate the precision of the graphical
representation of the result qualitatively in a way that is independent of the
regularization strength. For a quantitative documentation, the unfolding
results without explicit regularization should be published together with an
error matrix or its inverse. The widths of the bins of the corresponding histogram have to be
large enough to suppress excessive fluctuations.

A quantitative comparison of the R-L unfolding with other unfolding methods is
difficult, because in most of them a clear prescriptions for the choice of the
regularization strength is missing or doubtful. A sensible comparison
requires similar binning and regularization strengths in all methods. The
latter could be measured with the $p$-value. Independent of the unfolding method
that is used, in publications the values of $\chi^{2}$ obtained with and
without regularization should be given to indicate the regularization strength
and the reliability of the unfolded distribution.

Whenever it is possible to parametrize the true distribution, the parameters
should be fitted directly.

\section*{Acknowledgment}

I thank Gerhard Bohm for many valuable comments.

\section*{Appendix 1: The problem of the error assignment}%

\begin{figure}
[ptb]
\begin{center}
\includegraphics[
trim=0.000000in 0.099784in 0.000000in 0.000000in,
height=2.6083in,
width=5.9862in
]%
{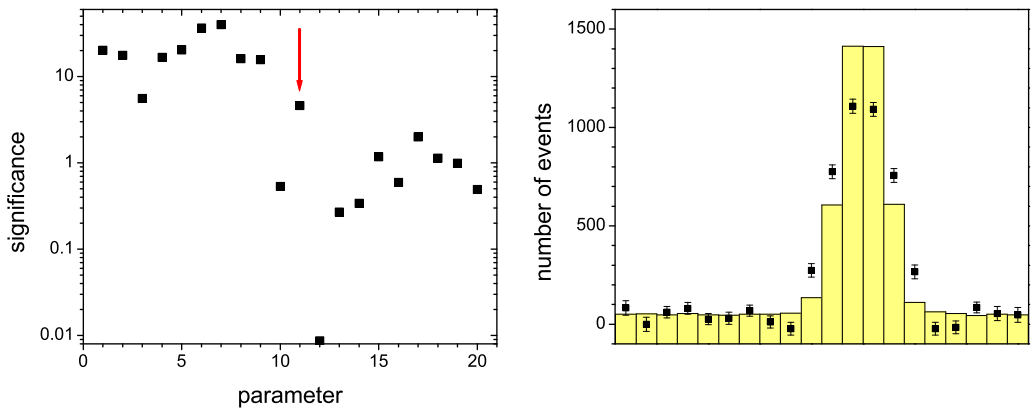}%
\caption{{\small Error assignment in regularized LSFs.}}%
\label{svderror}%
\end{center}
\end{figure}
In most unfolding schemes the oscillations are suppressed, either by
introduction of a penalty term in the fit, or by reduction of the effective
number of parameters \cite{blobel}. Both approaches constrain the fit and thus
reduce the errors. As a consequence the assigned uncertainties do not
necessarily cover the true distribution. An example is shown in Fig.
\ref{svderror} right hand side. The parameters of the LS fit have been
orthogonalized with a singular value decomposition (SVD) \cite{hoecker}. The
left hand plot shows the significance of the parameters which is defined as
the ratio of the parameter and its error as assigned by the fit. The $20$
parameters are ordered with decreasing eigenvalues. A smooth cut is applied at
parameter $\varepsilon_{c}=11$. Contributions are then weighted by
$\phi(\varepsilon)=\varepsilon/(\varepsilon+\varepsilon_{c)}$. In this way
oscillations are suppressed that might be caused by an abrupt cut, similar to
Gibbs oscillations as observed with Fourier approximations
\cite{hoecker,blobel}. Obviously the number of $11$ effectively used parameters
is insufficient to reproduce the peak and the true distribution is excluded.
With the addition of further parameters oscillations start to develop. The
problem is especially severe with low event numbers. With $10$ times more
events the discrepancy between the true distribution and the unfolded one is
considerably reduced.

Regularization with a curvature penalty reduces the statistical errors even in
the limit where the resolution is perfect. The errors presented by an
experiment that suffers from a limited resolution may be smaller than those of
a corresponding experiment with an ideal detector where unfolding is not required.

\section*{Appendix 2: The documentation of the results}%

\begin{figure}
[ptb]
\begin{center}
\includegraphics[
height=2.559in,
width=6.0831in
]%
{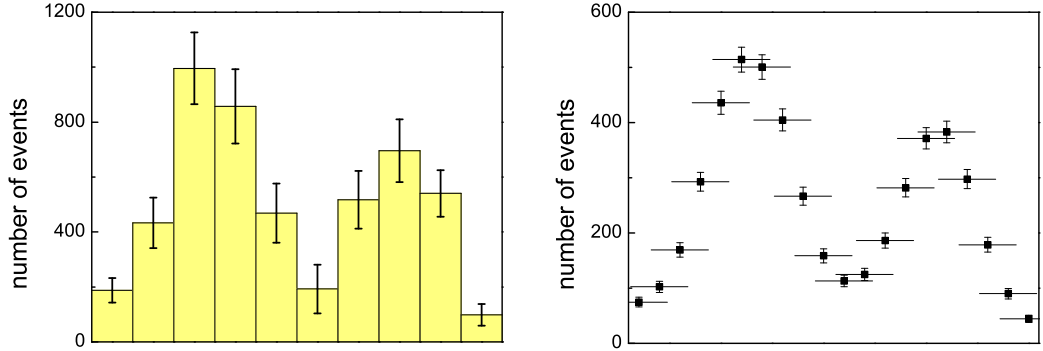}%
\caption{{\small Presentation of an unfolding result.}}%
\label{ilustrate}%
\end{center}
\end{figure}
In the following we present a possible way to document unfolding results such
that they can be compared to theoretical predictions and to other experiments.

The left hand plot of Fig. \ref{ilustrate} shows the result of a ML fit of the
content of the $10$ bins of a histogram without explicit regularization for
Example 1 with $5,000$ events. The errors are indicated. They are large due to
the strong negative correlation between adjacent bins which amounts to $80 \%$. The fitted values
together with the error matrix can be used for a quantitative comparison with
predictions. Instead of the error matrix its inverse could be presented. The inverse is in
fact the item that is required for parameter fitting. Even more information is
contained in the combination of the data vector and the response matrix. These
items, however, require some explanation to non-experts.

The right hand side of Fig. \ref{ilustrate} shows a possibility to indicate
the precision of an explicitly regularized unfolded histogram. The plot is
based on the same data as in the left hand plot. The vertical error bar
corresponds to the uncertainty of the bin content neglecting correlations
and the horizontal bars indicate the uncertainty in the location of the
events. In the absence of acceptance corrections the vertical error of bin $i$ is simply equal
to the square root of the bin content, $\sqrt{\theta_i}$. If the average acceptance
of the events in the bin is $\alpha_i$, the error is $\theta_i /\sqrt{\alpha_i\theta_i}$. 
The horizontal bar indicates the
experimental resolution. Such a graph is intended to show the likely shape of
the distribution but is not to be used for a quantitative comparison with
other results or predictions. It usually overestimates the uncertainties but for an
experienced scientist it indicates quite well the precision of a result.

\end{document}